\documentclass[aps,pra,twocolumn,nofootinbib,superscriptaddress]{revtex4-2}

\usepackage{amsmath,amssymb,bm}
\usepackage{bbm}
\usepackage{graphicx}

\newcommand{\I}{\mathbbm{1}}
\newcommand{\ket}[1]{\left|#1\right\rangle}
\newcommand{\bra}[1]{\left\langle#1\right|}
\newcommand{\proj}[1]{\ket{#1}\!\bra{#1}}
\newcommand{\Tr}{\operatorname{Tr}}
\newcommand{\FQ}{F_Q}

\begin{document}
	
	\title{Metrological Channels Induced by Encrypted Quantum Cloning:\\
		Optimal-Probe Fisher-Information Broadcasting and Single-Use Operational Recovery}
	
	\author{Nalan Kandirmaz}
	\email{nkandirmaz@mersin.edu.tr}
	\affiliation{Department of Physics, Mersin University, Mersin, T{\"u}rkiye}
	
	\date{July 25, 2026}
	
	\begin{abstract}
		Encrypted quantum cloning creates multiple individually inaccessible
		signal systems from which an unknown input state can be recovered by
		combining any selected signal with a shared quantum key. Here we study
		parameter estimation when a local sensing transformation is applied
		after encrypted cloning and before decryption. For the two-clone qubit
		protocol, a local Pauli rotation on one encrypted signal is shown to
		induce, after standard decryption, a parameter-dependent Pauli dephasing
		channel rather than a unitary rotation. We derive its Pauli-transfer
		representation and the corresponding quantum Fisher information (QFI) for
		arbitrary mixed qubit probes. We then distinguish raw QFI, which for an
		optimal decoded probe may
		remain simultaneously present in a decoded output and its complement,
		from decoder-accessible QFI, which is operationally single-use because a
		first decryption destroys the shared syndrome information needed for a
		second standard recovery. For $n\geq2$ encrypted clones, the
		effective channel family is independent of clone number up to a
		clone-parity-dependent permutation of the Pauli axes (the single-clone
		case is exceptional for $Y$-sensing, where the decoded map stays
		unitary). This produces a
		metrological parity switch: for fixed probe and generator, even and odd
		clone numbers can yield respectively vanishing and maximal QFI. Closed
		expressions are obtained for the degradation of this parity contrast
		under probe depolarization and Pauli-diagonal Bloch-vector contraction.
	\end{abstract}
	
	\maketitle
	
	\section{Introduction}
	
	The no-cloning theorem forbids the deterministic creation of independent
	perfect copies of an unknown quantum state~\cite{WoottersZurek1982}.
	Encrypted quantum cloning avoids this prohibition by producing multiple
	encrypted signal systems rather than simultaneously accessible copies. In
	the protocol introduced by Yamaguchi and Kempf, any number of encrypted
	signal qubits can be generated unitarily, while the initial state can be
	recovered from any selected signal only by consuming a shared key formed
	by noise qubits~\cite{YamaguchiKempf2026}. Each signal qubit is locally
	maximally mixed, and a successful decryption of one signal prevents a
	second independent recovery.
	
	Encrypted cloning has rapidly become an active subject. The protocol has
	been demonstrated experimentally on superconducting
	hardware~\cite{YamaguchiExp2026} and generalized to arbitrary
	dimensions~\cite{Ceara2026}. Its confidentiality has been analyzed in
	detail: unauthorized subsets of the storage register can retain a residual
	dependence on the input state confined to the $y$-Bloch component,
	governed by a parity-dependent leakage pattern~\cite{Gianini2026a,Gianini2026b},
	with a qudit extension~\cite{Bai2026}. The protocol has also been recast
	in access-structure terms, as a key-centered substructure of a
	quantum-secret-sharing scheme~\cite{Gianini2026c,LimLo2026}. These works
	concern the passive informativeness or recoverability of undecoded
	subsets. Here we address a complementary,
	metrological question: the effective channel and Fisher information
	produced when a parameter is imprinted after encoding and the state is
	then decoded.
	
	The metrological behavior of this construction, however, does not follow
	directly from its state-recovery properties. If
	a parameter is imprinted locally on one encrypted signal before
	decryption, three distinct questions arise: (i) what effective channel
	acts on the recovered probe, (ii) can the parameter information be
	localized by the shared key, and (iii) is the resulting Fisher
	information itself single-use? The last question requires care because
	QFI is not a conserved additive resource and does not obey a general
	monogamy relation of the form
	$F_Q(A)+F_Q(B)\leq F_Q(AB)$~\cite{Helstrom1976,Holevo1982,BraunsteinCaves1994,Paris2009,GLM2011}. Indeed,
	quantum Fisher information can be simultaneously present in several
	marginals of a state for suitable parameterized families, even though the
	underlying noncommuting states cannot be broadcast~\cite{Barnum1996,Lu2013}.
	
	Our contribution is not this broadcastability itself, but its operational
	realization within encrypted cloning: standard recovery converts a local
	Pauli rotation into a Pauli dephasing channel, and, for suitable probes,
	broadcast raw QFI should be distinguished from the QFI that is actually
	recoverable through the shared decoder. A useful operational distinction introduced here is
	\begin{equation*}
		\begin{aligned}
			&\text{raw QFI may be broadcast for suitable probes, whereas}\\
			&\text{decoder-accessible QFI is single-use.}
		\end{aligned}
	\end{equation*}
	
	The Yamaguchi--Kempf protocol establishes encrypted cloning as a
	state-recovery primitive~\cite{YamaguchiKempf2026}; we take that fixed
	protocol as given and ask a metrological question about it. To the best of
	our knowledge, previous studies have not analyzed the effective
	metrological channel generated by
	post-encoding parameter imprinting, nor the distinction between broadcast
	raw Fisher information and operationally recoverable Fisher information
	under the standard encrypted-cloning decoder. Our specific contributions
	--- the decoded effective channel, its quantum Fisher information, the
	raw/decoder-accessible distinction, and the clone-parity-controlled
	metrological switch --- are not addressed in that work or its subsequent
	confidentiality, qudit, and secret-sharing extensions.
	
	\section{Encrypted-cloning protocol}
	
	Let $A$ denote the unknown input qubit and let $(S_i,N_i)$,
	$i=1,\dots,n$, be signal--noise pairs initially prepared in Bell states
	\begin{equation}
		\ket{\phi_0}_{S_iN_i}=\frac{\ket{00}_{S_iN_i}+\ket{11}_{S_iN_i}}{\sqrt{2}}.
	\end{equation}
	We use
	\begin{equation}
		\ket{\phi_\mu}_{S_iN_i}=\left(\sigma_\mu^{(S_i)}\otimes\I^{(N_i)}\right)\ket{\phi_0}_{S_iN_i},
		\quad \mu=0,1,2,3,
	\end{equation}
	with $\sigma_0=\I$, $\sigma_1=X$, $\sigma_2=Y$, and $\sigma_3=Z$.
	
\begin{figure*}[t]
		\centering
		\includegraphics[width=\textwidth]{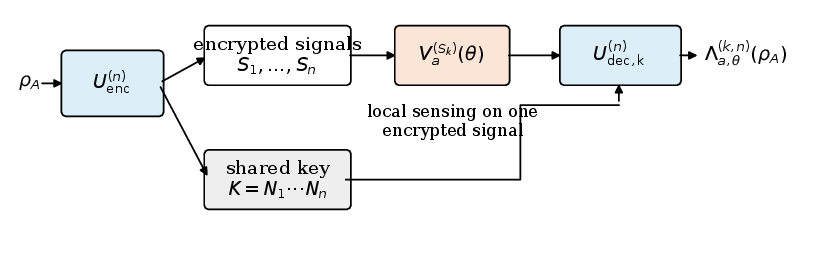}
		\caption{Post-encoding metrology with encrypted quantum cloning. The
			parameter is imprinted locally on a selected encrypted signal $S_k$.
			Standard recovery combines that signal with the shared key
			$K=N_1\cdots N_n$, producing the effective decoded channel
			$\Lambda_{a,\theta}^{(k,n)}$. Because the decoder acts on all noise
			qubits, the key is a shared, nonlocal recovery resource.}
		\label{fig:circuit}
	\end{figure*}

	The encrypted-cloning unitary is
	\begin{equation}
		\begin{split}
			U_{\mathrm{enc}}^{(n)}={}&\exp\!\left[-\frac{\pi i}{4}X^{(A)}\!\otimes\!\bigotimes_{i=1}^{n}X^{(S_i)}\right]\\
			&\times\exp\!\left[-\frac{\pi i}{4}Z^{(A)}\!\otimes\!\bigotimes_{i=1}^{n}Z^{(S_i)}\right].
		\end{split}
	\end{equation}
	For a pure input $\ket{\psi}_A$, the encoded state can be written as
	\begin{equation}
		\ket{\Psi_{\mathrm{enc}}}
		=\frac{1}{2}\sum_{\mu=0}^{3}\alpha_\mu^{-1}\,\sigma_\mu^{(A)}\ket{\psi}_A
		\bigotimes_{i=1}^{n}\ket{\phi_\mu}_{S_iN_i},
		\label{eq:encodedBell}
	\end{equation}
	where
	\begin{equation}
		\alpha_0=1,\qquad \alpha_1=\alpha_3=i,\qquad \alpha_2=-\bigl(i^{\,n+1}\bigr).
		\label{eq:alpha}
	\end{equation}
	To recover the state from a selected signal $S_k$, the decoder is
	\begin{equation}
		U_{\mathrm{dec},k}^{(n)}
		=\sum_{\mu=0}^{3}\alpha_\mu\proj{\phi_\mu}_{S_kN_k}
		\otimes\bigotimes_{\substack{\ell=1\\\ell\neq k}}^{n}
		\left(\sigma_\mu^{(N_\ell)}\right)^{T}.
		\label{eq:Udec}
	\end{equation}
	This expression makes the shared-key structure explicit: the decoder of
	one signal acts on all noise qubits (Fig.~\ref{fig:circuit}).
	
	The resulting access structure is closely related to quantum secret
	sharing: the encrypted signals constitute individually unauthorized
	shares, whereas recovery requires one selected signal together with the
	shared quantum key~\cite{Hillery1999,Cleve1999}. This correspondence has
	recently been made systematic, with encrypted cloning read as a
	key-centered substructure of a quantum-secret-sharing access structure in
	which a family of qualified sets shares a common non-qualified
	intersection that plays the role of the key~\cite{Gianini2026c,LimLo2026}.
	Within that classification the present protocol is ramp-like rather than
	perfectly threshold, since undecoded subsets may retain partial
	information. The objective here is nevertheless different from
	conventional secret reconstruction, because the present protocol realizes
	encrypted cloning with a consumable recovery resource, and because our
	concern below is metrological rather than access-structural.
	
	\section{Post-encoding sensing for two encrypted clones}
	
	After encoding, a local Pauli rotation is applied to $S_1$,
	\begin{equation}
		V_a^{(S_1)}(\theta)=\exp\!\left(-\frac{i\theta}{2}\sigma_a^{(S_1)}\right),
		\qquad a\in\{x,y,z\},
	\end{equation}
	with associated channel
	$\mathcal{V}_{a,\theta}^{(S_1)}(\rho)=V_a^{(S_1)}(\theta)\,\rho\,V_a^{(S_1)\dagger}(\theta)$.
	The effective decoded channel is
	\begin{equation}
		\Lambda_{a,\theta}^{(1,2)}
		=\mathcal{D}_1^{(2)}\circ\mathcal{V}_{a,\theta}^{(S_1)}\circ\mathcal{E}_2,
	\end{equation}
	where the encoding channel includes the initialized Bell-pair ancillas,
	\begin{equation*}
		\mathcal{E}_2(\rho_A)=U_{\mathrm{enc}}^{(2)}\!\left[\rho_A\otimes\proj{\phi_0}_{S_1N_1}\otimes\proj{\phi_0}_{S_2N_2}\right]U_{\mathrm{enc}}^{(2)\dagger},
	\end{equation*}
	and the decoder channel is, at first use,
	\begin{equation*}
		\mathcal{D}_k^{(n)}(\sigma)=\Tr_{\overline{S_k}}\!\left[
		U_{\mathrm{dec},k}^{(n)}\,\sigma\,U_{\mathrm{dec},k}^{(n)\dagger}\right],
	\end{equation*}
	where $\overline{S_k}$ denotes all registers other than $S_k$; this
	definition is restated in Eq.~\eqref{eq:deckdef} below for later use. Here the superscript
	$(1,2)$ records the sensed/recovered signal index and the clone number;
	Sec.~\ref{sec:generaln} generalizes this to $\Lambda_{a,\theta}^{(k,n)}$
	for a selected signal $k$ and clone number $n$.
	Writing
	\begin{equation}
		V_a(\theta)=\cos\frac{\theta}{2}\,\I-i\sin\frac{\theta}{2}\,\sigma_a,
	\end{equation}
	the identity and Pauli branches become correlated with orthogonal
	Bell-syndrome sectors after decryption. Explicitly, specializing the
	state of Appendix~\ref{app:singleuse}, Step~1
	[Eq.~\eqref{eq:branchreset}] to $n=2$, $k=1$, the state immediately
	after sensing and the first decoder is
	\begin{align}
		\ket{\Psi'}=\;&\tfrac12\cos\tfrac{\theta}{2}\sum_\mu(\sigma_\mu\ket{\psi})_A
		\ket{\phi_\mu}_{S_1N_1}\ket{\phi_0}_{S_2N_2}\nonumber\\
		&-\tfrac{i}{2}\sin\tfrac{\theta}{2}\sum_\mu\gamma_\mu(\sigma_\mu\ket{\psi})_A
		\ket{\phi_{a\cdot\mu}}_{S_1N_1}\ket{\phi_a}_{S_2N_2}.
	\end{align}
	where, for $n=2$,
	\begin{equation*}
		\gamma_\mu=\alpha_\mu^{-1}\alpha_{a\cdot\mu}\lambda_{a\mu}\lambda_{\mu,\,a\cdot\mu}.
	\end{equation*}
	Keeping these branch phases and the Bell-label permutation explicitly gives the
	recovered Pauli actions $Z$, $X$, and $Y$ for $a=x,y,z$, respectively, as
	displayed in Eqs.~\eqref{eq:Lambda2x}--\eqref{eq:Lambda2z}. The unused pair $S_2N_2$ is left in
	$\ket{\phi_0}$ by the $\cos$ branch
	and in $\ket{\phi_a}$ ($a\neq0$) by the $\sin$ branch; since
	$\langle\phi_0|\phi_a\rangle=0$, tracing out $A$, $N_1$, and $S_2N_2$
	kills the $\cos$--$\sin$ cross term exactly. Tracing out the unused
	systems therefore removes their cross terms and produces the
	random-unitary channels
	\begin{align}
		\Lambda_{x,\theta}^{(1,2)}(\rho)&=\cos^2\frac{\theta}{2}\,\rho+\sin^2\frac{\theta}{2}\,Z\rho Z,\label{eq:Lambda2x}\\
		\Lambda_{y,\theta}^{(1,2)}(\rho)&=\cos^2\frac{\theta}{2}\,\rho+\sin^2\frac{\theta}{2}\,X\rho X,\label{eq:Lambda2y}\\
		\Lambda_{z,\theta}^{(1,2)}(\rho)&=\cos^2\frac{\theta}{2}\,\rho+\sin^2\frac{\theta}{2}\,Y\rho Y.\label{eq:Lambda2z}
	\end{align}
	Defining
	\begin{equation}
		\tau_x=Z,\qquad \tau_y=X,\qquad \tau_z=Y,
	\end{equation}
	we obtain
	\begin{equation}
		\Lambda_{a,\theta}^{(1,2)}(\rho)=q_\theta\,\rho+(1-q_\theta)\,\tau_a\rho\tau_a,
		\qquad q_\theta=\cos^2\frac{\theta}{2}.
	\end{equation}
	In the ordered Pauli basis $(\I,X,Y,Z)$, the Pauli-transfer matrix
	$T_a(\theta)$ of $\Lambda_{a,\theta}^{(1,2)}$ is defined by
	$\Tr[\sigma_\mu\Lambda_{a,\theta}^{(1,2)}(\sigma_\nu)]=2\,[T_a(\theta)]_{\mu\nu}$,
	so that, for $\rho=(\I+\bm r\cdot\bm\sigma)/2$ and
	$\Lambda(\rho)=(\I+\bm s\cdot\bm\sigma)/2$,
	\begin{equation*}
		\begin{pmatrix}1\\\bm s\end{pmatrix}
		=T_a(\theta)\begin{pmatrix}1\\\bm r\end{pmatrix}.
	\end{equation*}
	Explicitly,
	\begin{align}
		T_x(\theta)&=\mathrm{diag}(1,\cos\theta,\cos\theta,1),\\
		T_y(\theta)&=\mathrm{diag}(1,1,\cos\theta,\cos\theta),\\
		T_z(\theta)&=\mathrm{diag}(1,\cos\theta,1,\cos\theta).
	\end{align}
	The subscript $a$ labels the \emph{sensing} axis rather than the
	resulting dephasing axis $\tau_a$; e.g.\ $T_x(\theta)$ leaves the $\I$-
	and $Z$-components invariant because $\tau_x=Z$.
	
	\section{Quantum Fisher information of the decoded channel}
	
	Let
	\begin{equation}
		\rho_0=\frac{1}{2}\left(\I+\bm{r}\cdot\bm{\sigma}\right),\qquad |\bm{r}|\leq1.
	\end{equation}
	Let $\bm{u}_a$ be the Bloch axis associated with $\tau_a$, and write
	\begin{equation}
		\bm{r}=m\bm{u}_a+\bm{r}_\perp,\qquad m=\bm{r}\cdot\bm{u}_a,\qquad
		\bm{r}_\perp\cdot\bm{u}_a=0.
	\end{equation}
	The output Bloch vector is
	\begin{equation}
		\bm{s}(\theta)=m\bm{u}_a+\cos\theta\,\bm{r}_\perp.
	\end{equation}
	Using the qubit Bloch-vector QFI formula~\cite{Zhong2013}
	\begin{equation}
		F_Q=|\partial_\theta\bm{s}|^2+\frac{(\bm{s}\cdot\partial_\theta\bm{s})^2}{1-|\bm{s}|^2},
	\end{equation}
	we obtain
	\begin{equation}
		F_Q^{(a)}=\frac{|\bm{r}_\perp|^2(1-m^2)\sin^2\theta}
		{1-m^2-|\bm{r}_\perp|^2\cos^2\theta}.
		\label{eq:QFIgeneral}
	\end{equation}
	For pure probes, $|\bm{r}|=1$, and therefore
	\begin{equation}
		F_Q^{(a)}=1-m^2=1-\langle\tau_a\rangle_\psi^2.
	\end{equation}
	In particular,
	\begin{equation}
		F_Q^{(x)}=1-r_z^2,\qquad F_Q^{(y)}=1-r_x^2,\qquad F_Q^{(z)}=1-r_y^2.
	\end{equation}
	These expressions hold on the regular interval $0<\theta<\pi$ for pure
	probes. At the endpoints $\theta=0,\pi$ the output rank changes for such
	pure-probe cases (where $|\bm{r}|=1$) and the SLD-based QFI can be
	discontinuous~\cite{Safranek2017}; there we adopt the continuous (Bures)
	extension, and the standard regularity assumptions underlying the
	Cram\'er--Rao bound require separate care. For generic mixed probes the
	output rank need not change at $\theta=0,\pi$; nonetheless
	Eq.~\eqref{eq:QFIgeneral} must be read with care wherever its denominator
	vanishes. In particular, at $m^2=1$ the pure probe is aligned with
	$\bm{u}_a$, so $\bm{r}_\perp=0$ and the output state is independent of
	$\theta$; therefore $F_Q=0$ by direct evaluation, rather than by inserting
	the resulting $0/0$ expression into Eq.~\eqref{eq:QFIgeneral}. All
	operational statements below are framed at the working point
	$\theta=\pi/2$, where these subtleties do not arise. For every finite
	visibility $0<\nu<1$, $\theta=\pi/2$ is the unique interior maximizer of
	the parity contrast $\Delta F_Y(\nu,\theta)$. At ideal visibility
	$\nu=1$, the contrast is instead constant on the regular interval
	$0<\theta<\pi$. Thus $\theta=\pi/2$ is used as a robust working point
	when finite visibility is included.
	
	\section{Optimal-probe broadcasting of raw QFI in the encrypted-cloning dilation}
	
	A Stinespring dilation of the decoded channel~\cite{NielsenChuang2000} is
	\begin{equation}
		W_{a,\theta}\ket{\psi}=\cos\frac{\theta}{2}\ket{\psi}\ket{0}_E
		+\sin\frac{\theta}{2}\,\tau_a\ket{\psi}\ket{1}_E.
	\end{equation}
	The environment state is
	\begin{equation}
		\rho_E(\theta)=
		\begin{pmatrix}
			\cos^2(\theta/2) & \tfrac{1}{2}\sin\theta\,\langle\tau_a\rangle\\[2mm]
			\tfrac{1}{2}\sin\theta\,\langle\tau_a\rangle & \sin^2(\theta/2)
		\end{pmatrix}.
	\end{equation}
	For pure probes, direct evaluation gives $F_Q[\rho_E(\theta)]=1$ for the
	minimal two-dimensional dilation above. In the sense of Ref.~\cite{Lu2013},
	the parameter is broadcast when a single $\theta$-independent channel
	$\mathcal{D}$ maps a parameterized input family to a bipartite output
	family whose two marginals each retain the input QFI. Here $\mathcal{D}$ is
	the encrypted-cloning decoder, which does not depend on $\theta$. Because
	the parameter is imprinted on the encrypted signal $S_k$ rather than on
	the bare probe, the parametric input to $\mathcal{D}$ is the post-encoding,
	post-sensing state
	$\sigma_\theta=\mathcal{V}^{(S_k)}_{a,\theta}\!\left(\mathcal{E}_n(\rho_A)\right)$;
	since the sensed signal is locally maximally mixed before the imprinting,
	its global pure-state QFI is probe-independent,
	$F_Q^{\mathrm{global}}=1$. The decoded output marginal carries
	$F_Q^{\mathrm{out}}=1-\langle\tau_a\rangle^2$.
	
	To establish the corresponding result directly for the \emph{physical}
	complementary subsystem, without assuming a $\theta$-independent isometry
	between Stinespring dilations, let $C_k=\overline{S_k}$ denote all
	registers except the recovered signal. For $n\geq2$, choose any unused pair
	$S_jN_j$ with $j\neq k$. From the explicit post-decoding state in
	Eq.~\eqref{eq:branchreset}, its reduced state is
	\begin{equation*}
		\rho_{S_jN_j}(\theta)=\cos^2\frac{\theta}{2}\proj{\phi_0}
		+\sin^2\frac{\theta}{2}\proj{\phi_a}.
	\end{equation*}
	Because $\ket{\phi_0}$ and $\ket{\phi_a}$ are orthogonal, the QFI of
	this pair is purely classical and, on $0<\theta<\pi$,
	\begin{equation*}
		F_Q[\rho_{S_jN_j}]
		=\frac{\sin^2\theta}{4}
		\left(\frac{1}{\cos^2(\theta/2)}+\frac{1}{\sin^2(\theta/2)}\right)=1.
	\end{equation*}
	Since $S_jN_j$ is obtained from $C_k$ by partial trace, while $C_k$ is
	itself a marginal of the global pure state, monotonicity of QFI gives
	\begin{equation*}
		1=F_Q[\rho_{S_jN_j}]\leq F_Q[\rho_{C_k}]\leq
		F_Q^{\mathrm{global}}=1,
	\end{equation*}
	and hence $F_Q[\rho_{C_k}]=1$. Thus, for an optimal decoded probe with
	$\langle\tau_a\rangle=0$, the decoded output and the physical complement
	both attain the global value,
	\begin{equation}
		F_Q^{\mathrm{out}}=F_Q^{\mathrm{comp}}=F_Q^{\mathrm{global}}=1,
	\end{equation}
	realizing the broadcasting criterion; equivalently, this is simultaneous
	marginal retention of QFI. Raw QFI therefore does not obey a general
	monogamy inequality under reduction, in contrast to the fact that
	noncommuting states themselves cannot be broadcast~\cite{Barnum1996}.
	
	The coexistence of
	$\FQ^{\mathrm{out}}=\FQ^{\mathrm{comp}}=\FQ^{\mathrm{global}}=1$
	does not violate the data-processing inequality. For each reduced channel
	separately, monotonicity gives
	$\FQ^{\mathrm{out}}\leq\FQ^{\mathrm{global}}$ and
	$\FQ^{\mathrm{comp}}\leq\FQ^{\mathrm{global}}$, and both bounds are
	saturated here. The quantity
	$\FQ^{\mathrm{out}}+\FQ^{\mathrm{comp}}$ is a sum over two different
	marginal channels and is not itself bounded by the data-processing
	inequality.
	
	We emphasize that this broadcasting does not constitute an information
	leak in the cryptographic sense. The complementary subsystem here contains
	the shared key and is therefore not equivalent to an unauthorized signal
	holder; the presence of QFI in this complement does not imply locally
	accessible information in an encrypted signal alone. An unauthorized holder
	of a signal alone has zero local QFI because its marginal is maximally
	mixed, as shown in Eqs.~\eqref{eq:signalmixed} and~\eqref{eq:signalqfi}.
	The single-use property of Sec.~\ref{sec:singleuse} is operationally
	different: after one standard recovery, a second decoder acting on another
	signal and the post-consumption key produces a maximally mixed output.
	
	\section{Decoder-accessible QFI and operational single use}
	\label{sec:singleuse}
	
	We define the decoder channel by tracing out every register except the
	recovered signal,
	\begin{equation}
		\mathcal{D}_k^{(n)}(\rho)=\Tr_{\overline{S_k}}\!\left[
		U_{\mathrm{dec},k}^{(n)}\,\rho\,U_{\mathrm{dec},k}^{(n)\dagger}\right],
		\label{eq:deckdef}
	\end{equation}
	where $\overline{S_k}$ denotes all registers other than $S_k$. The
	decoder-accessible QFI of a signal $S_j$ is then
	\begin{equation}
		F_{\mathrm{dec}}^{(j)}=F_Q\!\left[\mathcal{D}_j^{(n)}(\rho_{S_jK})\right],
		\qquad K=N_1\cdots N_n,
		\label{eq:Fdecdef}
	\end{equation}
	and, after the $k$th signal has been decoded,
	\begin{equation}
		F_{\mathrm{dec}}^{(j|k)}=F_Q\!\left[\mathcal{D}_j^{(n)}(\rho_{S_jK}^{\mathrm{after}\,k})\right].
	\end{equation}
	The parameter is imprinted only on the sensed signal $S_k$. Hence for
	$j=k$ the recovered channel is $\theta$-dependent. The single-use statement
	proved below concerns the ordered protocol in which $S_k$ is decoded first;
	a subsequent standard decoding of any $S_j$ with $j\neq k$ then carries no
	residual $\theta$-dependence.
	Before access to the shared key, every individual encrypted signal is
	maximally mixed~\cite{YamaguchiKempf2026}, and the local sensing --- a
	unitary on $S_k$ --- preserves this marginal,
	\begin{equation}
		\rho_{S_j}(\theta)=
		\begin{cases}
			V_{a,\theta}\,\tfrac{\I}{2}\,V_{a,\theta}^{\dagger}=\tfrac{\I}{2}, & j=k,\\[1mm]
			\tfrac{\I}{2}, & j\neq k.
		\end{cases}
		\label{eq:signalmixed}
	\end{equation}
	Consequently, its local QFI vanishes identically,
	\begin{equation}
		F_Q[\rho_{S_j}(\theta)]=0.
		\label{eq:signalqfi}
	\end{equation}
	Thus no parameter information is locally accessible from an unauthorized
	signal, even though raw QFI may remain encoded in correlations with the
	global complementary subsystem, which includes the shared key.
	
	Let $\Phi_b=\proj{\phi_b}$. The decoder acts on a Bell sector of $S_jN_j$
	as
	\begin{equation}
		\begin{split}
			U_{\mathrm{dec},j}^{(n)}\ket{\phi_b}_{S_jN_j}\otimes\ket{\omega}
			&=\alpha_b\ket{\phi_b}_{S_jN_j}\otimes W_b\ket{\omega},\\
			W_b&=\!\bigotimes_{\ell\neq j}\!\left(\sigma_b^{(N_\ell)}\right)^{T}\!,
		\end{split}
		\label{eq:decact}
	\end{equation}
	so on the off-diagonal blocks it multiplies by $\alpha_b\alpha_c^{*}$ and
	conjugates the key by $W_b(\cdot)W_c^{\dagger}$. Here $W_b$ is the
	correction the decoder $U_{\mathrm{dec},j}^{(n)}$ of
	Eq.~\eqref{eq:Udec} applies to the remaining key qubits $N_\ell$
	($\ell\neq j$) once the $S_jN_j$ sector is projected onto $\ket{\phi_b}$;
	the transpose follows from the Bell-state ricochet identity
	$(M\otimes\I)\ket{\phi_0}=(\I\otimes M^T)\ket{\phi_0}$, which transfers
	the Pauli correction from the signal side of a Bell pair to its noise
	qubit and gives the $(\sigma_\mu^{(N_\ell)})^T$ factors in the decoder
	unitary~\eqref{eq:Udec}.
	This correction operator is the mechanism by which single-use recovery
	is enforced: it is what the phase-sum argument of
	Appendix~\ref{app:singleuse} must overcome for a second decoder to
	extract any residual $\theta$-dependence.
	
	A key point is that the second decoder $\mathcal{D}_j^{(n)}$ acts on the
	\emph{entire} key $K=N_1\cdots N_n$, not only on $N_j$, so its $S_j$
	output is fixed by the full post-decryption state $\rho_{S_jK}^{(k)}$
	rather than by the pair marginal alone. Expanding this state in the Bell
	basis of $S_jN_j$,
	\begin{equation}
		\rho_{S_jK}^{(k)}=\sum_{b,c}\ket{\phi_b}\!\bra{\phi_c}_{S_jN_j}\otimes\Omega_{bc},
		\qquad \sum_{b}\Tr\Omega_{bb}=1,
		\label{eq:fullkeystate}
	\end{equation}
	with $\Omega_{bc}$ operators on $K\setminus N_j$, the pair marginal
	$\rho_{S_jN_j}^{(k)}=\cos^2(\theta/2)\Phi_0+\sin^2(\theta/2)\Phi_a$ retains
	only the diagonal blocks; the off-diagonal blocks $\Omega_{bc}$ ($b\neq
	c$) are generally nonzero and must be shown not to contribute.
	
	\emph{Lemma.} For every post-decryption state of the
	form~\eqref{eq:fullkeystate} produced by the first recovery --- i.e.\ with
	the unused pairs reset to definite Bell sectors, so that $\Omega_{bc}$ is
	supported on $b,c\in\{0,a\}$ (Appendix~\ref{app:singleuse}, Step~1) --- the
	standard decoder yields the maximally mixed signal,
	\begin{equation}
		\Tr_{K}\!\left[U_{\mathrm{dec},j}^{(n)}\,\rho_{S_jK}^{(k)}\,
		U_{\mathrm{dec},j}^{(n)\dagger}\right]=\frac{\I_{S_j}}{2}.
		\label{eq:lemma}
	\end{equation}
	
	\emph{Proof.} By~\eqref{eq:decact}, the decoded state is
	$\sum_{b,c}\alpha_b\alpha_c^{*}\ket{\phi_b}\!\bra{\phi_c}\otimes
	W_b\Omega_{bc}W_c^{\dagger}$. Using $\ket{\phi_\mu}=(\sigma_\mu\otimes\I)\ket{\phi_0}$
	one has $\Tr_{N_j}\ket{\phi_b}\!\bra{\phi_c}=\tfrac12\sigma_b\sigma_c$, so
	tracing $N_j$ and the remaining key gives
	\begin{equation}
		\begin{split}
			\rho_{S_j}^{\mathrm{out}}
			&=\frac12\sum_{b,c}\alpha_b\alpha_c^{*}\,\sigma_b\sigma_c\,g_{bc},\\
			g_{bc}&=\Tr_{K\setminus N_j}\!\left[W_b\Omega_{bc}W_c^{\dagger}\right].
		\end{split}
	\end{equation}
	The diagonal terms give
	$\tfrac12\sum_b|\alpha_b|^2\sigma_b^{2}\,\Tr\Omega_{bb}
	=\tfrac12\I\sum_b\Tr\Omega_{bb}=\tfrac12\I$,
	using $|\alpha_b|=1$, $\sigma_b^{2}=\I$, and the normalization
	in~\eqref{eq:fullkeystate}. For the off-diagonal terms, cyclicity gives
	$g_{bc}=\Tr[(W_c^{\dagger}W_b)\,\Omega_{bc}]$, in which $W_c^{\dagger}W_b$
	is, up to a phase, a nontrivial Pauli string on $K\setminus N_j$ whenever
	$b\neq c$. As proved in Appendix~\ref{app:singleuse}, the first decryption
	resets the unused pairs to definite Bell sectors, and each off-diagonal
	block $\Omega_{bc}$ then has vanishing overlap with $W_c^{\dagger}W_b$, so
	$g_{bc}=0$ for $b\neq c$. Only the diagonal survives, which
	proves~\eqref{eq:lemma}. $\square$
	
	The vanishing $g_{bc}=0$ for $b\neq c$ holds for every sensing axis and
	every clone number $n\geq2$; the general-$n$ proof, which reduces each
	off-diagonal contribution to a phase sum that cancels, is given in
	Appendix~\ref{app:singleuse}. Let $\rho^{\mathrm{after}\,k}(\rho_A,\theta)$ denote the full state after
	encoding the input probe $\rho_A$, sensing $S_k$, and applying the first
	decoder to $S_k$. The effective second-decoder channel is defined by
	\begin{equation*}
		\Gamma_{j|k,a,\theta}^{(n)}(\rho_A)
		\equiv\mathcal{D}_j^{(n)}\!\left[\rho^{\mathrm{after}\,k}(\rho_A,\theta)\right],
		\qquad j\neq k.
	\end{equation*}
	This is the same channel entering $F_{\mathrm{dec}}^{(j|k)}$ above. We consequently obtain
	\begin{equation}
		\Gamma_{j|k,a,\theta}^{(n)}(\rho_A)=\frac{\I}{2},\qquad
		F_{\mathrm{dec}}^{(j|k)}=0,\qquad j\neq k,
	\end{equation}
	for every clone number $n\geq2$, using the notation of
	Eq.~\eqref{eq:Fdecdef}. This proof assumes a specific ordering: the
	sensed signal $S_k$ is decoded first, so that the state of
	Eq.~\eqref{eq:branchreset} is the input to the second decoder acting on
	$S_j$. The reverse ordering, in which an unsensed signal is decoded
	before $S_k$, is not treated here or in Appendix~\ref{app:singleuse}.
	This is an operational statement about the standard decoder under that
	ordering, not a claim that all raw parameter information disappears
	from the global complement.
	
	Two points fix the operational meaning. First, the decoder-accessible QFI
	of Eq.~\eqref{eq:Fdecdef}, which for $j=k$ coincides with the
	decoded-output QFI of Eq.~\eqref{eq:QFIgeneral} because
	$\mathcal{D}_k^{(n)}(\rho_{S_kK})=\Lambda_{a,\theta}^{(k,n)}(\rho)$, is
	the QFI per independent
	encode--sense--decode protocol run; it sets the quantum Cram\'er--Rao
	bound $\mathrm{Var}(\hat\theta)\geq 1/(M\,F_Q)$ for $M$ such runs and for
	locally unbiased estimators under the usual regularity assumptions. It is
	locally attained by a projective measurement in the
	symmetric-logarithmic-derivative eigenbasis; since this eigenbasis
	generally depends on the unknown $\theta$, attainability is understood in
	the local (adaptive) sense, around a known operating point. The special
	value of the fixed-measurement construction of
	Sec.~\ref{sec:opmeas} is precisely that it removes this dependence for
	$\ket{+x}$, $Y$-sensing and odd $n\geq3$. Second, ``single-use'' refers to
	the consumption of the shared key within one recovery: estimating an
	unknown $\theta$ still proceeds by independent repetitions of the full
	encode--sense--decode protocol, each preparing and consuming its own key.
	The single-use property therefore constrains what can be recovered from a
	single encoded instance, not the asymptotic estimation itself.
	
	\section{General-$n$ channel and clone-parity control}
	\label{sec:generaln}
	
	For $n\geq2$, tracing over the unused encoded pairs removes the
	interbranch coherences and the post-decoding channel remains in the
	random-unitary family
	\begin{equation}
		\Lambda_{a,\theta}^{(k,n)}(\rho)
		=\cos^2\frac{\theta}{2}\,\rho+\sin^2\frac{\theta}{2}\,\tau_a^{(n)}\rho\,\tau_a^{(n)}.
	\end{equation}
	The weights are independent of $n$, while the effective Pauli axis
	depends on clone-number parity through the full $n$-dependent decoder
	phase structure derived in Appendix~\ref{app:parity}. For even $n$,
	\begin{equation}
		\begin{array}{c|ccc}
			a & x & y & z\\\hline
			\tau_a^{(n)} & Z & X & Y
		\end{array},
		\label{eq:evenTable}
	\end{equation}
	whereas for odd $n$,
	\begin{equation}
		\begin{array}{c|ccc}
			a & x & y & z\\\hline
			\tau_a^{(n)} & Z & Y & X
		\end{array}.
		\label{eq:oddTable}
	\end{equation}
	The single-clone case $n=1$ requires separate treatment, because there
	are then no unused signal pairs to trace over. For $X$- and $Z$-sensing
	the decoded map still reduces to the odd-$n$ dephasing channels, with
	axes $\tau_x^{(1)}=Z$ and $\tau_z^{(1)}=X$. For $Y$-sensing, however, no
	interbranch coherence is removed and the decoded map remains the unitary
	rotation
	\begin{equation}
		\Lambda_{y,\theta}^{(1,1)}(\rho)=V_y(\theta)\,\rho\,V_y^\dagger(\theta),
		\label{eq:n1unitary}
	\end{equation}
	rather than a $Y$-dephasing channel. Accordingly, the random-unitary
	channel~\eqref{eq:evenTable}--\eqref{eq:oddTable} and the noise formulas
	of Sec.~\ref{sec:noise} are stated for $n\geq2$, and the odd-$n$
	metrological switch below refers to odd $n\geq3$; the $n=1$ $Y$-branch is
	the unitary special case~\eqref{eq:n1unitary}, whose metrological
	consequences are noted where relevant.
	Thus the effective channel family is clone-number independent up to a
	parity-controlled permutation of the Pauli axes. The complete Bell-label
	derivation leading to the parity tables is given in
	Appendix~\ref{app:parity}. This dependence is a physical effect rather
	than a choice of labeling: for a fixed input probe, a fixed laboratory
	generator, and the fixed decoder convention, the recovered-output QFI
	itself differs between even and odd clone number, as quantified in
	Sec.~\ref{sec:noise} and therefore directly measurable. It shares an algebraic origin in the clone-number-dependent branch phases
	with the parity-dependent confidentiality leakage of undecoded subsets~\cite{Gianini2026a,Gianini2026b}, which in
	the access-structure picture marks those subsets as ramp-like
	intermediate sets rather than forbidden ones~\cite{Gianini2026c}. The two
	effects are nonetheless operationally distinct: the leakage is a residual
	dependence in a passive reduced state, whereas the present effect is a
	measurable Fisher-information switch in the decoded metrological channel.
	We emphasize that this switch is a property of a \emph{fixed} probe and
	sensing generator: if the probe is instead re-optimized independently
	for each parity, the maximal pure-state QFI remains $F_Q=1$ in both
	cases (Eq.~\eqref{eq:QFIgeneral} with $m=0$), so the parity dependence
	controls the response of a fixed sensing configuration rather than the
	optimized metrological capacity of the channel.
	
	\section{Parity switching under probe depolarization and Pauli-diagonal noise}
	\label{sec:noise}
	
	Let a pure probe with Bloch vector $\bm{r}_0$ be depolarized
	as~\cite{DemkowiczDobrzanski2012}
	\begin{equation}
		\rho_\nu=\nu\rho_0+(1-\nu)\frac{\I}{2},\qquad 0\leq\nu\leq1.
	\end{equation}
	Defining $m_0=\bm{r}_0\cdot\bm{u}_a^{(n)}$, Eq.~\eqref{eq:QFIgeneral} gives
	\begin{equation}
		F_\nu^{(a,n)}=\frac{\nu^2(1-m_0^2)(1-\nu^2m_0^2)\sin^2\theta}
		{1-\nu^2m_0^2-\nu^2(1-m_0^2)\cos^2\theta}.
	\end{equation}
	For $m_0=0$,
	\begin{equation}
		F_{\nu,\mathrm{opt}}=\frac{\nu^2\sin^2\theta}{1-\nu^2\cos^2\theta}.
	\end{equation}
	At $\theta=\pi/2$, this reduces to $F_{\nu,\mathrm{opt}}=\nu^2$.
	
\begin{figure}[t]
		\centering
		\includegraphics[width=\columnwidth]{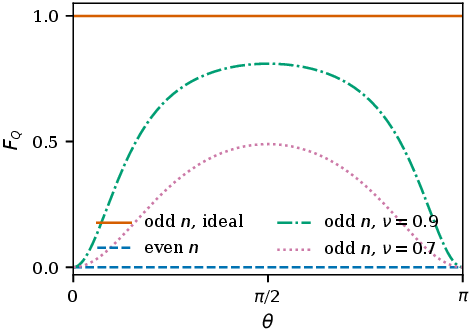}
		\caption{Parity-resolved QFI for a $\ket{+x}$ probe under $Y$-sensing.
			Odd clone numbers yield $F_Q=1$ at ideal visibility, whereas even clone
			numbers yield $F_Q=0$. Finite-visibility curves ($\nu=0.9,0.7$) show
			the depolarized odd-$n\geq3$ response. Because the even-parity QFI
			vanishes for this fixed probe and generator, the odd-parity curve also
			equals the parity contrast $\Delta F_Y(\nu,\theta)$. The ideal
			$\nu=1$ curve applies to all odd $n$, including the unitary $n=1$
			$Y$-branch. At the rank-changing endpoints $\theta=0,\pi$ the ideal
			curve is shown using the continuous (Bures) extension of the QFI.}
		\label{fig:switch}
	\end{figure}

\begin{figure}[t]
		\centering
		\includegraphics[width=\columnwidth]{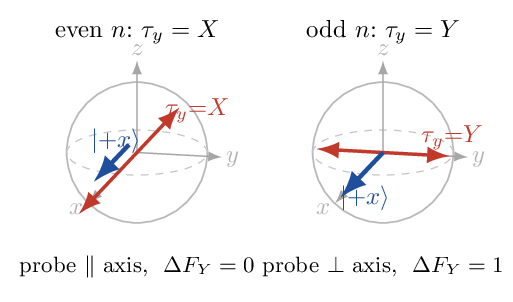}
		\caption{Bloch-sphere schematic of the parity switch for $Y$-sensing of a
			$\ket{+x}$ probe. For even clone number the effective dephasing axis is
			$\tau_y=X$, parallel to the probe, so the induced decoded $X$-dephasing
			channel leaves the recovered probe
			invariant and $\Delta F_Y=0$; for odd clone number the axis is
			$\tau_y=Y$, perpendicular to the probe, giving maximal contrast
			$\Delta F_Y=1$.}
		\label{fig:generator}
	\end{figure}

	For the fixed probe $\ket{+x}$ and $Y$-sensing (Fig.~\ref{fig:generator}),
	even $n$ gives $\tau_y^{(\mathrm{even})}=X$, so the probe is parallel to
	the effective dephasing axis and
	\begin{equation}
		F_{\mathrm{even}}^{(Y,+x)}=0.
	\end{equation}
	Odd $n\geq3$ gives $\tau_y^{(\mathrm{odd})}=Y$, and therefore
	\begin{equation}
		F_{\mathrm{odd}}^{(Y,+x)}=\frac{\nu^2\sin^2\theta}{1-\nu^2\cos^2\theta}.
	\end{equation}
	For the exceptional single-clone case, $Y$-sensing instead produces the
	unitary channel~\eqref{eq:n1unitary}, whose QFI for the depolarized
	$\ket{+x}$ probe is
	\begin{equation}
		F^{(Y,+x)}_{n=1}=\nu^2,
	\end{equation}
	independent of $\theta$ and equal to the odd-$n\geq3$ value only at the
	working point $\theta=\pi/2$. In particular, at ideal visibility $\nu=1$
	all odd $n$ (including $n=1$) yield unit QFI, so the parity switch is
	already exhibited by the $n=1$ versus $n=2$ comparison; the distinction
	between $n=1$ and odd $n\geq3$ appears only away from $\theta=\pi/2$ under
	finite visibility.
	The parity contrast is
	\begin{equation}
		\Delta F_Y(\nu,\theta)=\frac{\nu^2\sin^2\theta}{1-\nu^2\cos^2\theta}.
	\end{equation}
	At $\theta=\pi/2$,
	\begin{equation}
		\Delta F_Y=\nu^2.
	\end{equation}
	In the ideal case, $F_{\mathrm{even}}=0$ and $F_{\mathrm{odd}}=1$
	(Fig.~\ref{fig:switch}).
	
	For $Z$-sensing with the same probe,
	\begin{equation}
		\tau_z^{(\mathrm{even})}=Y,\qquad \tau_z^{(\mathrm{odd})}=X,
	\end{equation}
	so the switch direction is reversed. For $X$-sensing,
	\begin{equation}
		\tau_x^{(\mathrm{even})}=\tau_x^{(\mathrm{odd})}=Z,
	\end{equation}
	and no parity contrast occurs. The ideal maximal gaps are therefore
	\begin{equation}
		\Delta F_X^{\max}=0,\qquad \Delta F_Y^{\max}=1,\qquad \Delta F_Z^{\max}=1.
	\end{equation}
	For a Pauli-diagonal Bloch contraction,
	\begin{equation}
		(r_x,r_y,r_z)\longmapsto(\eta_xr_x,\eta_yr_y,\eta_zr_z),
	\end{equation}
	which is a completely positive unital qubit channel only when the
	$\eta_i$ satisfy the Fujiwara--Algoet conditions
	$1\pm\eta_z\geq|\eta_x\pm\eta_y|$ (ordinary $Z$-dephasing being the special
	case $\eta_x=\eta_y=\eta$, $\eta_z=1$), the $Y$-sensing contrast for
	$\ket{+x}$ becomes
	\begin{equation}
		\Delta F_Y=\frac{\eta_x^2\sin^2\theta}{1-\eta_x^2\cos^2\theta}.
	\end{equation}
	We use this map only as a phenomenological contraction model; when a
	specific $(\eta_x,\eta_y,\eta_z)$ is intended physically, the above CP
	conditions must be imposed.
	
	\section{Operational measurement of the parity switch}
	\label{sec:opmeas}
	
	The parity switch is accessible with a single fixed measurement, without
	any optimization over the probe or the POVM. Take the probe $\ket{+x}$ and
	$Y$-sensing. For odd clone number $n\geq3$, $\tau_y^{(\mathrm{odd})}=Y$, so the
	decoded state is
	\begin{equation}
		\rho_{\mathrm{odd}}(\theta)=\tfrac{1}{2}\left(\I+\cos\theta\,X\right),
	\end{equation}
	and a projective measurement in the $X$ basis gives outcome probabilities
	\begin{equation}
		p_\pm(\theta)=\frac{1\pm\cos\theta}{2},
	\end{equation}
	with $\partial_\theta p_\pm=\mp\tfrac12\sin\theta$ and hence
	\begin{align}
		F_C^{\mathrm{odd}}
		&=\sum_{\pm}\frac{[\partial_\theta p_\pm(\theta)]^2}{p_\pm(\theta)}\nonumber\\
		&=\frac{\sin^2\theta}{4}\left(\frac{1}{p_+}+\frac{1}{p_-}\right)
		=1=F_Q^{\mathrm{odd}}
	\end{align}
	on the regular interval $0<\theta<\pi$. For even clone number,
	$\tau_y^{(\mathrm{even})}=X$, and $\ket{+x}$ is a fixed point of the
	induced $X$-dephasing, so $\rho_{\mathrm{even}}(\theta)=\proj{+x}$ is
	$\theta$-independent and
	\begin{equation}
		F_C^{\mathrm{even}}=0=F_Q^{\mathrm{even}}.
	\end{equation}
	The same probe- and parameter-independent $X$-basis measurement therefore
	realizes the full $0\leftrightarrow1$ switch. Under depolarizing
	visibility $\nu$, the decoded state becomes
	$\rho_{\mathrm{odd},\nu}(\theta)=\tfrac{1}{2}(\I+\nu\cos\theta\,X)$, and the
	same measurement yields
	\begin{equation}
		F_C^{\mathrm{odd}}(\nu,\theta)=\frac{\nu^2\sin^2\theta}{1-\nu^2\cos^2\theta},
	\end{equation}
	which coincides with the decoder-accessible QFI. The optimal measurement
	is thus a single fixed projective measurement on the decoded qubit, and
	the parity contrast is directly visible in its outcome statistics rather
	than only in an optimized information measure. This provides an
	experimentally testable signature of the parity switch. For the
	single-clone case the same $X$-basis outcome statistics apply, since the
	unitary $n=1$ $Y$-branch yields the identical probabilities
	$p_\pm=(1\pm\nu\cos\theta)/2$; however, its decoder-accessible QFI is
	$\nu^2$, so at finite visibility the fixed $X$-basis measurement is no
	longer optimal for $n=1$ and saturates the bound only at $\theta=\pi/2$.
	
	\section{Illustrative visibility model}
	
\begin{figure}[t]
		\centering
		\includegraphics[width=\columnwidth]{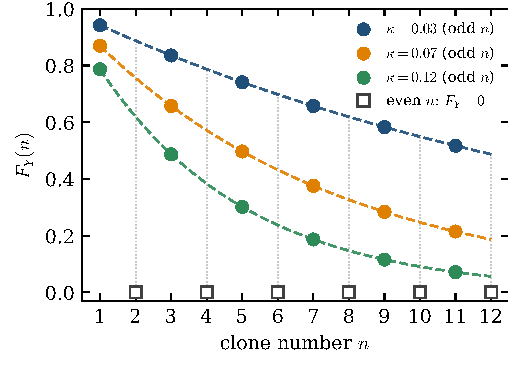}
		\caption{Illustrative odd-parity visibility envelope. The parity-resolved
			contrast $F_Y(n)$ vanishes identically at even clone number and follows
			the phenomenological envelope $e^{-2\kappa n}$ at odd $n$ (integer
			markers); the decay rate is set by the effective implementation parameter
			$\kappa$. The model is not calibrated to a specific device.}
		\label{fig:hardware}
	\end{figure}

	We illustrate how the parity contrast degrades with implementation
	imperfection using a phenomenological model; it is not
	calibrated to a specific device, and a quantitative link would require
	measured gate-error data for a concrete encrypted-cloning circuit. At the
	optimal working point,
	\begin{equation}
		\Delta F(\pi/2)=\nu^2.
	\end{equation}
	For a minimum observable contrast $F_{\min}$,
	\begin{equation}
		\nu\geq\sqrt{F_{\min}}.
	\end{equation}
	A simple clone-number-dependent visibility model is
	\begin{equation}
		\nu(n)=e^{-\kappa n}.
	\end{equation}
	Because the contrast vanishes identically at even clone number, the
	observed response is parity-resolved,
	\begin{equation}
		F_Y(n)=
		\begin{cases}
			0, & n\ \text{even},\\[1mm]
			e^{-2\kappa n}, & n\ \text{odd},
		\end{cases}
	\end{equation}
	so $e^{-2\kappa n}$ is the odd-parity envelope rather than a smooth
	function of $n$ (Fig.~\ref{fig:hardware}). The largest odd clone number
	retaining a contrast above
	$F_{\min}$ is the largest odd integer not exceeding
	\begin{equation}
		\left\lfloor -\frac{\ln F_{\min}}{2\kappa}\right\rfloor .
	\end{equation}
	A gate-level model may instead use
	\begin{equation}
		\nu(n)\simeq(1-\epsilon_2)^{G_2(n)}(1-\epsilon_1)^{G_1(n)}.
	\end{equation}
	
	\section{Continuous-variable outlook}
	
	As an outlook, and without proof, we note the ingredients that a
	continuous-variable realization would require: a bosonic encrypted-cloning
	or secret-sharing encoder, local post-encoding sensing, a finite-resource
	shared key, and a selected decoder. A natural optical resource is the
	two-mode squeezed vacuum
	\begin{equation}
		\begin{split}
			\ket{\mathrm{TMSV}(r_K)}&=\sqrt{1-\lambda^2}\sum_{m=0}^{\infty}\lambda^m\ket{m,m},\\
			\lambda&=\tanh r_K.
		\end{split}
	\end{equation}
	The relevant finite-resource quantities are
	\begin{equation}
		F_{\mathrm{auth}}(r_K),\qquad F_{\mathrm{comp}}(r_K),\qquad F_{\mathrm{dec}}^{(j|k)}(r_K),
	\end{equation}
	for coherent, squeezed, and displaced-squeezed probes. A full
	continuous-variable treatment requires an explicit bosonic encoder and
	decoder and is left for future work.
	
	\section{Conclusions}
	
	Local Pauli sensing on an encrypted signal is converted by standard
	decryption into a parameter-dependent Pauli dephasing channel. For an
	optimal decoded probe ($\langle\tau_a\rangle=0$), its QFI
	can be maximal even when the same parameter information remains visible in
	the complementary subsystem, so the QFI of the two marginals need not
	satisfy an additive monogamy inequality; for a generic probe the
	complementary subsystem retains the full QFI while the decoded output
	generally does not, so this simultaneous retention is probe-dependent.
	In contrast, decoder-accessible
	QFI is operationally single-use: after the sensed signal $S_k$ is
	decoded first, every second
	standard decoder acting on another signal $j\neq k$ produces a maximally
	mixed output; the reverse ordering is not addressed here. The mechanism is that
	the first decryption resets the unused pairs to definite Bell sectors,
	each with maximally mixed signal marginal, while the residual interbranch
	coherences drop out of the second decoder.
	
	For $n\geq2$ clones, the channel probabilities are unchanged while
	the decoder phase produces an even--odd permutation of the effective
	Pauli axis; the single-clone case is exceptional for $Y$-sensing, where
	the decoded map remains unitary. This generates a metrological parity
	switch, with ideal maximal QFI contrast equal to one for $Y$- and
	$Z$-sensing and zero for $X$-sensing. The mixed-probe and noise formulas
	show how this contrast is reduced under probe depolarization and
	Pauli-diagonal Bloch-vector contraction. These results show that encrypted
	quantum cloning can realize a parity-controlled metrological channel whose decoded
	response depends on the clone-number parity. This response is specific to
	the encrypted-cloning decoder and its clone-parity-dependent phase
	structure, and differs from the usual fidelity- or QFI-distribution
	problems studied in approximate cloning
	machines~\cite{Song2013,Xiao2014,Yao2014}. The resulting even--odd switch,
	directly visible in a fixed-measurement protocol, provides an
	experimentally testable signature.
	
	\appendix
	
	\section{Bell-label derivation of the parity tables}
	\label{app:parity}
	
	This appendix derives the parity-dependent Pauli tables used in the main
	text directly from the encoded state, Eqs.~\eqref{eq:encodedBell}--\eqref{eq:Udec}.
	The derivation is based on the Bell-label permutation induced by left
	multiplication of a Pauli operator on the sensed signal.
	
	The Bell labels are permuted by left multiplication of the Pauli
	operator acting on the sensed signal,
	\begin{equation}
		\sigma_a^{(S_k)}\ket{\phi_\mu}_{S_kN_k}=\eta_{a\mu}\ket{\phi_{\pi_a(\mu)}}_{S_kN_k},
		\label{eq:BellPerm}
	\end{equation}
	where the permutations and phases follow directly from the Pauli
	multiplication rule. In the convention used here they are
	\begin{equation}
		\begin{array}{c|cccc}
			& \mu=0 & \mu=1 & \mu=2 & \mu=3\\\hline
			X & (1,+1) & (0,+1) & (3,+i) & (2,-i)\\
			Y & (2,+1) & (3,-i) & (0,+1) & (1,+i)\\
			Z & (3,+1) & (2,+i) & (1,-i) & (0,+1)
		\end{array},
		\label{eq:BellActionTable}
	\end{equation}
	where each entry denotes $(\pi_a(\mu),\eta_{a\mu})$.
	
	Substitution of Eq.~\eqref{eq:BellPerm} into the encoded
	state~\eqref{eq:encodedBell}, followed by the decoder~\eqref{eq:Udec},
	produces the branch factor
	\begin{equation}
		\gamma_{a\mu}^{(n)}=\alpha_\mu^{-1}\alpha_{\pi_a(\mu)}\,\eta_{a\mu}\,
		\bigl(\lambda_{\mu,\,\pi_a(\mu)}\bigr)^{n-1},
	\end{equation}
	which includes the phase picked up on the $n-1$ unused signal--noise
	pairs and coincides, with $\pi_a(\mu)=a\cdot\mu$ and
	$\eta_{a\mu}=\lambda_{a\mu}$, with $\gamma_\mu$ of
	Appendix~\ref{app:singleuse}, Step~1. This extra factor is generally
	branch dependent and therefore must be retained; it cannot be absorbed
	into a single global phase. Combining the four $\mu$ branches with the
	complete phase $\gamma_{a\mu}^{(n)}$ gives the recovered Pauli actions
	\begin{equation}
		\begin{array}{c|ccc}
			& a=x & a=y & a=z\\\hline
			n\ \mathrm{even} & Z & X & Y\\
			n\ \mathrm{odd} & Z & Y & X
		\end{array}.
	\end{equation}
	Thus the $x$ generator is parity invariant, whereas the $y$ and $z$
	generators are interchanged when the clone number changes parity. The
	parity dependence comes from the combined $n$ dependence of
	$\alpha_\mu$ and $\bigl(\lambda_{\mu,\pi_a(\mu)}\bigr)^{n-1}$;
	direct evaluation of the four branches for even and odd $n$ yields the
	two-cycle in Eqs.~\eqref{eq:evenTable} and~\eqref{eq:oddTable} without
	discarding any branch-dependent phase. We stress that this two-cycle fixes
	the effective Pauli \emph{axis} $\tau_a^{(n)}$ for every $n$, including
	$n=1$; whether the decoded map is a $\tau_a$-dephasing channel or a
	$\tau_a$-unitary additionally requires that at least one unused encoded
	pair be traced out, which occurs only for $n\geq2$. This is why the
	single-clone $Y$-branch, whose axis is nonetheless correctly $Y$, remains
	the unitary rotation~\eqref{eq:n1unitary}.
	
	\section{General-$n$ proof of the single-use property}
	\label{app:singleuse}
	
	Here we prove the Lemma of Sec.~\ref{sec:singleuse} for arbitrary $n\geq2$,
	i.e.\ $g_{bc}=0$ for $b\neq c$ in Eq.~\eqref{eq:lemma}. We use
	$\sigma_\mu\sigma_\nu=\lambda_{\mu\nu}\sigma_{\mu\cdot\nu}$ with
	$|\lambda_{\mu\nu}|=1$, the Bell ricochet identity
	$(\I\otimes\sigma_\nu^{T})\ket{\phi_\mu}=\lambda_{\mu\nu}\ket{\phi_{\mu\cdot\nu}}$,
	the sensing action
	$V_a(\theta)\ket{\phi_\mu}=\cos\tfrac{\theta}{2}\ket{\phi_\mu}
	-i\sin\tfrac{\theta}{2}\lambda_{a\mu}\ket{\phi_{a\cdot\mu}}$,
	and $\sigma_\mu^{T}=\sigma_\mu^{*}=t_\mu\sigma_\mu$ with
	$t_0=t_1=t_3=1$, $t_2=-1$.
	
	\emph{Step 1 (branch reset).} Applying the sensing on $S_k$ and the first
	decoder $U_{\mathrm{dec},k}^{(n)}=\sum_\nu\alpha_\nu\Phi_\nu^{S_kN_k}
	\otimes\bigotimes_{\ell\neq k}(\sigma_\nu^{N_\ell})^{T}$ to the encoded
	state, the $S_kN_k$ projector selects $\nu=\mu$ from the $\cos$ term and
	$\nu=a\cdot\mu$ from the $\sin$ term, while the corrections act on each
	unused pair by the ricochet identity. Since $\mu\cdot\mu=0$ and
	$\mu\cdot(a\cdot\mu)=a$, every unused pair is reset to a definite Bell
	state,
	\begin{align}
		\ket{\Psi'}=\;&\tfrac12\cos\tfrac{\theta}{2}\sum_\mu(\sigma_\mu\ket{\psi})_A
		\ket{\phi_\mu}_{S_kN_k}\!\bigotimes_{\ell\neq k}\ket{\phi_0}_{S_\ell N_\ell}
		\nonumber\\
		&-\tfrac{i}{2}\sin\tfrac{\theta}{2}\sum_\mu\gamma_\mu(\sigma_\mu\ket{\psi})_A
		\ket{\phi_{a\cdot\mu}}_{S_kN_k}\!\bigotimes_{\ell\neq k}\ket{\phi_a}_{S_\ell N_\ell},
		\label{eq:branchreset}
	\end{align}
	with $\gamma_\mu=\alpha_\mu^{-1}\alpha_{a\cdot\mu}\lambda_{a\mu}
	(\lambda_{\mu,\,a\cdot\mu})^{\,n-1}$. In particular each unused pair has
	marginal $\cos^2\tfrac{\theta}{2}\Phi_0+\sin^2\tfrac{\theta}{2}\Phi_a$, so
	in~\eqref{eq:fullkeystate} only the blocks $b,c\in\{0,a\}$ occur.
	
	\emph{Step 2 (off-diagonal block).} Tracing out $A$, $S_k$ and
	$\{S_\ell\}_{\ell\neq j,k}$ and projecting onto
	$\bra{\phi_0}_{S_jN_j}\cdots\ket{\phi_a}_{S_jN_j}$ selects the
	$\cos$--$\sin$ cross term. With $\Tr_{S}\ket{\phi_\mu}\!\bra{\phi_\nu}
	=\tfrac12(\sigma_\nu\sigma_\mu)^{T}$ and $\Tr_A[\sigma_\mu\proj{\psi}\sigma_{\mu'}]
	=\bra{\psi}\sigma_{\mu'}\sigma_\mu\ket{\psi}$,
	\begin{equation}
		\begin{split}
			\Omega_{0a}={}&\frac{i\cos\tfrac{\theta}{2}\sin\tfrac{\theta}{2}}{4}
			\sum_{\mu,\mu'}\gamma_{\mu'}^{*}\bra{\psi}\sigma_{\mu'}\sigma_\mu\ket{\psi}\\
			&\times\tfrac12(\sigma_{a\cdot\mu'}\sigma_\mu)^{T}_{N_k}\!
			\bigotimes_{\ell\neq j,k}\tfrac12\sigma_a^{T}.
		\end{split}
	\end{equation}
	
	\emph{Step 3 (reduction).} With $b,c=0,a$ the second-decoder correction is
	$W_a^{\dagger}W_0=W_a^{\dagger}=t_a^{\,n-1}\sigma_a^{\otimes(n-1)}$ on
	$K\setminus N_j$, so $g_{0a}=t_a^{\,n-1}\Tr[(\bigotimes_{\ell\neq j}\sigma_a)\,\Omega_{0a}]$.
	Each factor $\ell\neq j,k$ gives $\Tr[\sigma_a\cdot\tfrac12\sigma_a^{T}]=t_a$,
	and the $N_k$ factor gives $\tfrac12\Tr[\sigma_a(\sigma_{a\cdot\mu'}\sigma_\mu)^{T}]
	=t_a\lambda_{a\mu'}^{-1}\delta_{\mu\mu'}$. The intermediate steps here use
	the transpose invariance of the trace, $\Tr[\sigma_a M^{T}]=\Tr[M\sigma_a^{T}]
	=t_a\Tr[M\sigma_a]$, followed by $\sigma_{a\cdot\mu'}=\lambda_{a\mu'}^{-1}\sigma_a\sigma_{\mu'}$
	and $\sigma_a^{2}=\I$:
	\begin{align}
		\Tr[\sigma_a(\sigma_{a\cdot\mu'}\sigma_\mu)^{T}]
		&=t_a\Tr[\sigma_{a\cdot\mu'}\sigma_\mu\sigma_a]\nonumber\\
		&=t_a\lambda_{a\mu'}^{-1}\Tr[\sigma_a^{2}\sigma_{\mu'}\sigma_\mu]\nonumber\\
		&=t_a\lambda_{a\mu'}^{-1}\Tr[\sigma_{\mu'}\sigma_\mu]
		=2t_a\lambda_{a\mu'}^{-1}\delta_{\mu\mu'}.
	\end{align}
	The $\ket{\psi}$ dependence collapses through $\delta_{\mu\mu'}$ and
	$\bra{\psi}\sigma_\mu^2\ket{\psi}=1$, leaving
	\begin{equation}
		g_{0a}=\frac{i}{4}\cos\tfrac{\theta}{2}\sin\tfrac{\theta}{2}\;S(n,a),
		\qquad
		S(n,a)=\sum_\mu \gamma_\mu^{*}\lambda_{a\mu}^{-1},
	\end{equation}
	where $t_a^{2(n-1)}=1$ was used. Thus $g_{0a}$ is independent of the input
	state and fixed by the phase sum $S(n,a)$.
	
	\emph{Step 4 (phase sum vanishes).} Writing
	$\lambda_{\mu,\,a\cdot\mu}=\lambda_{a\mu}^{-1}\chi_{\mu a}$ with
	$\sigma_\mu\sigma_a\sigma_\mu=\chi_{\mu a}\sigma_a$, $\chi_{\mu a}=\pm1$,
	the summand becomes
	\begin{equation}
		T_\mu\equiv\gamma_\mu^{*}\lambda_{a\mu}^{-1}
		=\alpha_\mu\,\alpha_{a\cdot\mu}^{-1}\,\lambda_{a\mu}^{\,n-3}\,\chi_{\mu a}^{\,n-1}.
	\end{equation}
	The map $\mu\mapsto a\cdot\mu$ is an involution, and using
	$\lambda_{a,\,a\cdot\mu}=\lambda_{a\mu}^{-1}$, $\chi_{a\cdot\mu,\,a}=\chi_{\mu a}$,
	one finds $T_\mu T_{a\cdot\mu}=\chi_{\mu a}^{2(n-1)}=1$, i.e.\
	$T_{a\cdot\mu}=T_\mu^{*}$. Each $T_\mu\in\{\pm1,\pm i\}$, and the relation
	$T_{a\cdot\mu}=T_\mu^{*}$ pairs the four labels into two involution orbits,
	each orbit contributing $T_\mu+T_{a\cdot\mu}=2\,\mathrm{Re}\,T_\mu$. For
	$a=X,Z$ the nonzero $T_\mu$ are purely imaginary for every $n$, so each
	orbit contributes zero. For $a=Y$ the $T_\mu$ are purely imaginary when
	$n$ is even (again giving zero per orbit) and real when $n$ is odd, in
	which case the two orbits carry the opposite values $+1$ and $-1$ and
	cancel against each other. Explicitly,
	\begin{equation}
		\begin{array}{c|c|cccc}
			n & a & T_0 & T_1 & T_2 & T_3\\ \hline
			\text{even} & X & -i & +i & +i & -i\\
			\text{even} & Y & -i & -i & +i & +i\\
			\text{even} & Z & -i & +i & -i & +i\\
			\text{odd}  & X & -i & +i & +i & -i\\
			\text{odd}  & Y & -1 & +1 & -1 & +1\\
			\text{odd}  & Z & -i & -i & +i & +i
		\end{array}
	\end{equation}
	the pattern repeating up to an overall sign as $n$ increases by two. In
	every case $S(n,a)=\sum_\mu T_\mu=0$ for every axis $a$ and every $n\geq2$. Hence
	$g_{0a}=0$, and by Hermiticity $g_{a0}=g_{0a}^{*}=0$; all other
	off-diagonal blocks vanish already in~\eqref{eq:fullkeystate}. Together
	with the diagonal contribution $\tfrac12\I$, this proves
	Eq.~\eqref{eq:lemma} for all $n\geq2$. $\square$
	
	We have independently confirmed every step of this derivation, and the
	identity $g_{bc}=0$ itself, by explicit matrix evaluation for clone
	numbers $2\leq n\leq5$ and all sensing axes. The NumPy script used for
	this independent encode--sense--decode verification is provided in
	the Supplemental Material~\cite{SupplementalMaterial}.
	
	\bibliographystyle{apsrev4-2}
	\bibliography{Referenceclon}

\begin{thebibliography}{26}%
\makeatletter
\providecommand \@ifxundefined [1]{%
 \@ifx{#1\undefined}
}%
\providecommand \@ifnum [1]{%
 \ifnum #1\expandafter \@firstoftwo
 \else \expandafter \@secondoftwo
 \fi
}%
\providecommand \@ifx [1]{%
 \ifx #1\expandafter \@firstoftwo
 \else \expandafter \@secondoftwo
 \fi
}%
\providecommand \natexlab [1]{#1}%
\providecommand \enquote  [1]{``#1''}%
\providecommand \bibnamefont  [1]{#1}%
\providecommand \bibfnamefont [1]{#1}%
\providecommand \citenamefont [1]{#1}%
\providecommand \href@noop [0]{\@secondoftwo}%
\providecommand \href [0]{\begingroup \@sanitize@url \@href}%
\providecommand \@href[1]{\@@startlink{#1}\@@href}%
\providecommand \@@href[1]{\endgroup#1\@@endlink}%
\providecommand \@sanitize@url [0]{\catcode `\\12\catcode `\$12\catcode
  `\&12\catcode `\#12\catcode `\^12\catcode `\_12\catcode `\%12\relax}%
\providecommand \@@startlink[1]{}%
\providecommand \@@endlink[0]{}%
\providecommand \url  [0]{\begingroup\@sanitize@url \@url }%
\providecommand \@url [1]{\endgroup\@href {#1}{\urlprefix }}%
\providecommand \urlprefix  [0]{URL }%
\providecommand \Eprint [0]{\href }%
\providecommand \doibase [0]{https://doi.org/}%
\providecommand \selectlanguage [0]{\@gobble}%
\providecommand \bibinfo  [0]{\@secondoftwo}%
\providecommand \bibfield  [0]{\@secondoftwo}%
\providecommand \translation [1]{[#1]}%
\providecommand \BibitemOpen [0]{}%
\providecommand \bibitemStop [0]{}%
\providecommand \bibitemNoStop [0]{.\EOS\space}%
\providecommand \EOS [0]{\spacefactor3000\relax}%
\providecommand \BibitemShut  [1]{\csname bibitem#1\endcsname}%
\let\auto@bib@innerbib\@empty
%</preamble>
\bibitem [{\citenamefont {Wootters}\ and\ \citenamefont
  {Zurek}(1982)}]{WoottersZurek1982}%
  \BibitemOpen
  \bibfield  {author} {\bibinfo {author} {\bibfnamefont {W.~K.}\ \bibnamefont
  {Wootters}}\ and\ \bibinfo {author} {\bibfnamefont {W.~H.}\ \bibnamefont
  {Zurek}},\ }\href@noop {} {\bibfield  {journal} {\bibinfo  {journal}
  {Nature}\ }\textbf {\bibinfo {volume} {299}},\ \bibinfo {pages} {802}
  (\bibinfo {year} {1982})}\BibitemShut {NoStop}%
\bibitem [{\citenamefont {Yamaguchi}\ and\ \citenamefont
  {Kempf}(2026)}]{YamaguchiKempf2026}%
  \BibitemOpen
  \bibfield  {author} {\bibinfo {author} {\bibfnamefont {K.}~\bibnamefont
  {Yamaguchi}}\ and\ \bibinfo {author} {\bibfnamefont {A.}~\bibnamefont
  {Kempf}},\ }\href@noop {} {\bibfield  {journal} {\bibinfo  {journal} {Phys.
  Rev. Lett.}\ }\textbf {\bibinfo {volume} {136}},\ \bibinfo {pages} {010801}
  (\bibinfo {year} {2026})}\BibitemShut {NoStop}%
\bibitem [{\citenamefont {Yamaguchi}\ \emph {et~al.}(2026)\citenamefont
  {Yamaguchi}, \citenamefont {Rullk{\"o}tter}, \citenamefont {Shehzad},
  \citenamefont {Wagner}, \citenamefont {Tutschku},\ and\ \citenamefont
  {Kempf}}]{YamaguchiExp2026}%
  \BibitemOpen
  \bibfield  {author} {\bibinfo {author} {\bibfnamefont {K.}~\bibnamefont
  {Yamaguchi}}, \bibinfo {author} {\bibfnamefont {L.}~\bibnamefont
  {Rullk{\"o}tter}}, \bibinfo {author} {\bibfnamefont {I.}~\bibnamefont
  {Shehzad}}, \bibinfo {author} {\bibfnamefont {S.~J.}\ \bibnamefont {Wagner}},
  \bibinfo {author} {\bibfnamefont {C.}~\bibnamefont {Tutschku}},\ and\
  \bibinfo {author} {\bibfnamefont {A.}~\bibnamefont {Kempf}},\ }\href@noop {}
  {\bibinfo {title} {Experimental demonstration that qubits can be cloned at
  will, if encrypted with a single-use decryption key}} (\bibinfo {year}
  {2026}),\ \Eprint {https://arxiv.org/abs/2602.10695} {arXiv:2602.10695}
  \BibitemShut {NoStop}%
\bibitem [{\citenamefont {Cear{\u{a}}}(2026)}]{Ceara2026}%
  \BibitemOpen
  \bibfield  {author} {\bibinfo {author} {\bibfnamefont {F.-I.}\ \bibnamefont
  {Cear{\u{a}}}},\ }\href {https://doi.org/10.1103/b32z-r5p7} {\bibfield
  {journal} {\bibinfo  {journal} {Phys. Rev. A}\ }\textbf {\bibinfo {volume}
  {114}},\ \bibinfo {pages} {022450} (\bibinfo {year} {2026})}\BibitemShut
  {NoStop}%
\bibitem [{\citenamefont {Gianini}\ \emph
  {et~al.}(2026{\natexlab{a}})\citenamefont {Gianini}, \citenamefont {Hasan},
  \citenamefont {Mio}, \citenamefont {Cimato},\ and\ \citenamefont
  {Damiani}}]{Gianini2026a}%
  \BibitemOpen
  \bibfield  {author} {\bibinfo {author} {\bibfnamefont {G.}~\bibnamefont
  {Gianini}}, \bibinfo {author} {\bibfnamefont {O.}~\bibnamefont {Hasan}},
  \bibinfo {author} {\bibfnamefont {C.}~\bibnamefont {Mio}}, \bibinfo {author}
  {\bibfnamefont {S.}~\bibnamefont {Cimato}},\ and\ \bibinfo {author}
  {\bibfnamefont {E.}~\bibnamefont {Damiani}},\ }\href@noop {} {\bibinfo
  {title} {Encrypted clones can leak: Classification of informative subsets in
  quantum encrypted cloning}} (\bibinfo {year} {2026}{\natexlab{a}}),\ \Eprint
  {https://arxiv.org/abs/2604.10155} {arXiv:2604.10155} \BibitemShut {NoStop}%
\bibitem [{\citenamefont {Gianini}\ \emph
  {et~al.}(2026{\natexlab{b}})\citenamefont {Gianini}, \citenamefont {Cimato},
  \citenamefont {Lin}, \citenamefont {Hasan},\ and\ \citenamefont
  {Damiani}}]{Gianini2026b}%
  \BibitemOpen
  \bibfield  {author} {\bibinfo {author} {\bibfnamefont {G.}~\bibnamefont
  {Gianini}}, \bibinfo {author} {\bibfnamefont {S.}~\bibnamefont {Cimato}},
  \bibinfo {author} {\bibfnamefont {J.}~\bibnamefont {Lin}}, \bibinfo {author}
  {\bibfnamefont {O.}~\bibnamefont {Hasan}},\ and\ \bibinfo {author}
  {\bibfnamefont {E.}~\bibnamefont {Damiani}},\ }\href@noop {} {\bibinfo
  {title} {Full characterization of informative subsets in quantum encrypted
  cloning}} (\bibinfo {year} {2026}{\natexlab{b}}),\ \Eprint
  {https://arxiv.org/abs/2605.27421} {arXiv:2605.27421} \BibitemShut {NoStop}%
\bibitem [{\citenamefont {Bai}\ \emph {et~al.}(2026)\citenamefont {Bai},
  \citenamefont {Zhou},\ and\ \citenamefont {Luo}}]{Bai2026}%
  \BibitemOpen
  \bibfield  {author} {\bibinfo {author} {\bibfnamefont {C.-M.}\ \bibnamefont
  {Bai}}, \bibinfo {author} {\bibfnamefont {X.-L.}\ \bibnamefont {Zhou}},\ and\
  \bibinfo {author} {\bibfnamefont {Y.}~\bibnamefont {Luo}},\ }\href@noop {}
  {\bibinfo {title} {Classification of informative subsets in quantum encrypted
  cloning on qudits}} (\bibinfo {year} {2026}),\ \Eprint
  {https://arxiv.org/abs/2605.11642} {arXiv:2605.11642} \BibitemShut {NoStop}%
\bibitem [{\citenamefont {Gianini}\ \emph
  {et~al.}(2026{\natexlab{c}})\citenamefont {Gianini}, \citenamefont {Cimato},
  \citenamefont {Lin}, \citenamefont {Hasan}, \citenamefont {Mio},\ and\
  \citenamefont {Damiani}}]{Gianini2026c}%
  \BibitemOpen
  \bibfield  {author} {\bibinfo {author} {\bibfnamefont {G.}~\bibnamefont
  {Gianini}}, \bibinfo {author} {\bibfnamefont {S.}~\bibnamefont {Cimato}},
  \bibinfo {author} {\bibfnamefont {J.}~\bibnamefont {Lin}}, \bibinfo {author}
  {\bibfnamefont {O.}~\bibnamefont {Hasan}}, \bibinfo {author} {\bibfnamefont
  {C.}~\bibnamefont {Mio}},\ and\ \bibinfo {author} {\bibfnamefont
  {E.}~\bibnamefont {Damiani}},\ }\href@noop {} {\bibinfo {title} {Beyond the
  canonical protocol: Quantum encrypted cloning from secret-sharing access
  structures}} (\bibinfo {year} {2026}{\natexlab{c}}),\ \Eprint
  {https://arxiv.org/abs/2606.06552} {arXiv:2606.06552} \BibitemShut {NoStop}%
\bibitem [{\citenamefont {Lim}\ and\ \citenamefont {Lo}(2026)}]{LimLo2026}%
  \BibitemOpen
  \bibfield  {author} {\bibinfo {author} {\bibfnamefont {Z.~L.}\ \bibnamefont
  {Lim}}\ and\ \bibinfo {author} {\bibfnamefont {H.-K.}\ \bibnamefont {Lo}},\
  }\href@noop {} {\bibinfo {title} {Encrypted cloning, absolute maximal
  entanglement and quantum secret sharing}} (\bibinfo {year} {2026}),\ \Eprint
  {https://arxiv.org/abs/2605.26866} {arXiv:2605.26866} \BibitemShut {NoStop}%
\bibitem [{\citenamefont {Helstrom}(1976)}]{Helstrom1976}%
  \BibitemOpen
  \bibfield  {author} {\bibinfo {author} {\bibfnamefont {C.~W.}\ \bibnamefont
  {Helstrom}},\ }\href@noop {} {\emph {\bibinfo {title} {Quantum Detection and
  Estimation Theory}}}\ (\bibinfo  {publisher} {Academic Press},\ \bibinfo
  {address} {New York},\ \bibinfo {year} {1976})\BibitemShut {NoStop}%
\bibitem [{\citenamefont {Holevo}(1982)}]{Holevo1982}%
  \BibitemOpen
  \bibfield  {author} {\bibinfo {author} {\bibfnamefont {A.~S.}\ \bibnamefont
  {Holevo}},\ }\href@noop {} {\emph {\bibinfo {title} {Probabilistic and
  Statistical Aspects of Quantum Theory}}}\ (\bibinfo  {publisher}
  {North-Holland},\ \bibinfo {address} {Amsterdam},\ \bibinfo {year}
  {1982})\BibitemShut {NoStop}%
\bibitem [{\citenamefont {Braunstein}\ and\ \citenamefont
  {Caves}(1994)}]{BraunsteinCaves1994}%
  \BibitemOpen
  \bibfield  {author} {\bibinfo {author} {\bibfnamefont {S.~L.}\ \bibnamefont
  {Braunstein}}\ and\ \bibinfo {author} {\bibfnamefont {C.~M.}\ \bibnamefont
  {Caves}},\ }\href@noop {} {\bibfield  {journal} {\bibinfo  {journal} {Phys.
  Rev. Lett.}\ }\textbf {\bibinfo {volume} {72}},\ \bibinfo {pages} {3439}
  (\bibinfo {year} {1994})}\BibitemShut {NoStop}%
\bibitem [{\citenamefont {Paris}(2009)}]{Paris2009}%
  \BibitemOpen
  \bibfield  {author} {\bibinfo {author} {\bibfnamefont {M.~G.}\ \bibnamefont
  {Paris}},\ }\href@noop {} {\bibfield  {journal} {\bibinfo  {journal} {Int. J.
  Quantum Inf.}\ }\textbf {\bibinfo {volume} {7}},\ \bibinfo {pages} {125}
  (\bibinfo {year} {2009})}\BibitemShut {NoStop}%
\bibitem [{\citenamefont {Giovannetti}\ \emph {et~al.}(2011)\citenamefont
  {Giovannetti}, \citenamefont {Lloyd},\ and\ \citenamefont
  {Maccone}}]{GLM2011}%
  \BibitemOpen
  \bibfield  {author} {\bibinfo {author} {\bibfnamefont {V.}~\bibnamefont
  {Giovannetti}}, \bibinfo {author} {\bibfnamefont {S.}~\bibnamefont {Lloyd}},\
  and\ \bibinfo {author} {\bibfnamefont {L.}~\bibnamefont {Maccone}},\
  }\href@noop {} {\bibfield  {journal} {\bibinfo  {journal} {Nat. Photonics}\
  }\textbf {\bibinfo {volume} {5}},\ \bibinfo {pages} {222} (\bibinfo {year}
  {2011})}\BibitemShut {NoStop}%
\bibitem [{\citenamefont {Barnum}\ \emph {et~al.}(1996)\citenamefont {Barnum},
  \citenamefont {Caves}, \citenamefont {Fuchs}, \citenamefont {Jozsa},\ and\
  \citenamefont {Schumacher}}]{Barnum1996}%
  \BibitemOpen
  \bibfield  {author} {\bibinfo {author} {\bibfnamefont {H.}~\bibnamefont
  {Barnum}}, \bibinfo {author} {\bibfnamefont {C.~M.}\ \bibnamefont {Caves}},
  \bibinfo {author} {\bibfnamefont {C.~A.}\ \bibnamefont {Fuchs}}, \bibinfo
  {author} {\bibfnamefont {R.}~\bibnamefont {Jozsa}},\ and\ \bibinfo {author}
  {\bibfnamefont {B.}~\bibnamefont {Schumacher}},\ }\href@noop {} {\bibfield
  {journal} {\bibinfo  {journal} {Phys. Rev. Lett.}\ }\textbf {\bibinfo
  {volume} {76}},\ \bibinfo {pages} {2818} (\bibinfo {year}
  {1996})}\BibitemShut {NoStop}%
\bibitem [{\citenamefont {Lu}\ \emph {et~al.}(2013)\citenamefont {Lu},
  \citenamefont {Sun}, \citenamefont {Wang}, \citenamefont {Luo},\ and\
  \citenamefont {Oh}}]{Lu2013}%
  \BibitemOpen
  \bibfield  {author} {\bibinfo {author} {\bibfnamefont {X.-M.}\ \bibnamefont
  {Lu}}, \bibinfo {author} {\bibfnamefont {Z.}~\bibnamefont {Sun}}, \bibinfo
  {author} {\bibfnamefont {X.}~\bibnamefont {Wang}}, \bibinfo {author}
  {\bibfnamefont {S.}~\bibnamefont {Luo}},\ and\ \bibinfo {author}
  {\bibfnamefont {C.}~\bibnamefont {Oh}},\ }\href@noop {} {\bibfield  {journal}
  {\bibinfo  {journal} {Phys. Rev. A}\ }\textbf {\bibinfo {volume} {87}},\
  \bibinfo {pages} {050302} (\bibinfo {year} {2013})}\BibitemShut {NoStop}%
\bibitem [{\citenamefont {Hillery}\ \emph {et~al.}(1999)\citenamefont
  {Hillery}, \citenamefont {Bu{\v{z}}ek},\ and\ \citenamefont
  {Berthiaume}}]{Hillery1999}%
  \BibitemOpen
  \bibfield  {author} {\bibinfo {author} {\bibfnamefont {M.}~\bibnamefont
  {Hillery}}, \bibinfo {author} {\bibfnamefont {V.}~\bibnamefont
  {Bu{\v{z}}ek}},\ and\ \bibinfo {author} {\bibfnamefont {A.}~\bibnamefont
  {Berthiaume}},\ }\href@noop {} {\bibfield  {journal} {\bibinfo  {journal}
  {Phys. Rev. A}\ }\textbf {\bibinfo {volume} {59}},\ \bibinfo {pages} {1829}
  (\bibinfo {year} {1999})}\BibitemShut {NoStop}%
\bibitem [{\citenamefont {Cleve}\ \emph {et~al.}(1999)\citenamefont {Cleve},
  \citenamefont {Gottesman},\ and\ \citenamefont {Lo}}]{Cleve1999}%
  \BibitemOpen
  \bibfield  {author} {\bibinfo {author} {\bibfnamefont {R.}~\bibnamefont
  {Cleve}}, \bibinfo {author} {\bibfnamefont {D.}~\bibnamefont {Gottesman}},\
  and\ \bibinfo {author} {\bibfnamefont {H.-K.}\ \bibnamefont {Lo}},\
  }\href@noop {} {\bibfield  {journal} {\bibinfo  {journal} {Phys. Rev. Lett.}\
  }\textbf {\bibinfo {volume} {83}},\ \bibinfo {pages} {648} (\bibinfo {year}
  {1999})}\BibitemShut {NoStop}%
\bibitem [{\citenamefont {Zhong}\ \emph {et~al.}(2013)\citenamefont {Zhong},
  \citenamefont {Sun}, \citenamefont {Ma}, \citenamefont {Wang},\ and\
  \citenamefont {Nori}}]{Zhong2013}%
  \BibitemOpen
  \bibfield  {author} {\bibinfo {author} {\bibfnamefont {W.}~\bibnamefont
  {Zhong}}, \bibinfo {author} {\bibfnamefont {Z.}~\bibnamefont {Sun}}, \bibinfo
  {author} {\bibfnamefont {J.}~\bibnamefont {Ma}}, \bibinfo {author}
  {\bibfnamefont {X.}~\bibnamefont {Wang}},\ and\ \bibinfo {author}
  {\bibfnamefont {F.}~\bibnamefont {Nori}},\ }\href@noop {} {\bibfield
  {journal} {\bibinfo  {journal} {Phys. Rev. A}\ }\textbf {\bibinfo {volume}
  {87}},\ \bibinfo {pages} {022337} (\bibinfo {year} {2013})}\BibitemShut
  {NoStop}%
\bibitem [{\citenamefont {{\v{S}}afr{\'a}nek}(2017)}]{Safranek2017}%
  \BibitemOpen
  \bibfield  {author} {\bibinfo {author} {\bibfnamefont {D.}~\bibnamefont
  {{\v{S}}afr{\'a}nek}},\ }\href@noop {} {\bibfield  {journal} {\bibinfo
  {journal} {Phys. Rev. A}\ }\textbf {\bibinfo {volume} {95}},\ \bibinfo
  {pages} {052320} (\bibinfo {year} {2017})}\BibitemShut {NoStop}%
\bibitem [{\citenamefont {Nielsen}\ and\ \citenamefont
  {Chuang}(2000)}]{NielsenChuang2000}%
  \BibitemOpen
  \bibfield  {author} {\bibinfo {author} {\bibfnamefont {M.~A.}\ \bibnamefont
  {Nielsen}}\ and\ \bibinfo {author} {\bibfnamefont {I.~L.}\ \bibnamefont
  {Chuang}},\ }\href@noop {} {\emph {\bibinfo {title} {Quantum Computation and
  Quantum Information}}}\ (\bibinfo  {publisher} {Cambridge University Press},\
  \bibinfo {address} {Cambridge},\ \bibinfo {year} {2000})\BibitemShut
  {NoStop}%
\bibitem [{\citenamefont {Demkowicz-Dobrza{\'n}ski}\ \emph
  {et~al.}(2012)\citenamefont {Demkowicz-Dobrza{\'n}ski}, \citenamefont
  {Ko{\l}ody{\'n}ski},\ and\ \citenamefont
  {Gu{\c{t}}{\u{a}}}}]{DemkowiczDobrzanski2012}%
  \BibitemOpen
  \bibfield  {author} {\bibinfo {author} {\bibfnamefont {R.}~\bibnamefont
  {Demkowicz-Dobrza{\'n}ski}}, \bibinfo {author} {\bibfnamefont
  {J.}~\bibnamefont {Ko{\l}ody{\'n}ski}},\ and\ \bibinfo {author}
  {\bibfnamefont {M.}~\bibnamefont {Gu{\c{t}}{\u{a}}}},\ }\href@noop {}
  {\bibfield  {journal} {\bibinfo  {journal} {Nat. Commun.}\ }\textbf {\bibinfo
  {volume} {3}},\ \bibinfo {pages} {1063} (\bibinfo {year} {2012})}\BibitemShut
  {NoStop}%
\bibitem [{\citenamefont {Song}\ \emph {et~al.}(2013)\citenamefont {Song},
  \citenamefont {Luo}, \citenamefont {Li},\ and\ \citenamefont
  {Chang}}]{Song2013}%
  \BibitemOpen
  \bibfield  {author} {\bibinfo {author} {\bibfnamefont {H.}~\bibnamefont
  {Song}}, \bibinfo {author} {\bibfnamefont {S.}~\bibnamefont {Luo}}, \bibinfo
  {author} {\bibfnamefont {N.}~\bibnamefont {Li}},\ and\ \bibinfo {author}
  {\bibfnamefont {L.}~\bibnamefont {Chang}},\ }\href@noop {} {\bibfield
  {journal} {\bibinfo  {journal} {Phys. Rev. A}\ }\textbf {\bibinfo {volume}
  {88}},\ \bibinfo {pages} {042121} (\bibinfo {year} {2013})}\BibitemShut
  {NoStop}%
\bibitem [{\citenamefont {Xiao}\ \emph {et~al.}(2014)\citenamefont {Xiao},
  \citenamefont {Yao}, \citenamefont {Zhou},\ and\ \citenamefont
  {Wang}}]{Xiao2014}%
  \BibitemOpen
  \bibfield  {author} {\bibinfo {author} {\bibfnamefont {X.}~\bibnamefont
  {Xiao}}, \bibinfo {author} {\bibfnamefont {Y.}~\bibnamefont {Yao}}, \bibinfo
  {author} {\bibfnamefont {L.-M.}\ \bibnamefont {Zhou}},\ and\ \bibinfo
  {author} {\bibfnamefont {X.}~\bibnamefont {Wang}},\ }\href@noop {} {\bibfield
   {journal} {\bibinfo  {journal} {Sci. Rep.}\ }\textbf {\bibinfo {volume}
  {4}},\ \bibinfo {pages} {7361} (\bibinfo {year} {2014})}\BibitemShut
  {NoStop}%
\bibitem [{\citenamefont {Yao}\ \emph {et~al.}(2014)\citenamefont {Yao},
  \citenamefont {Ge}, \citenamefont {Xiao}, \citenamefont {Wang},\ and\
  \citenamefont {Sun}}]{Yao2014}%
  \BibitemOpen
  \bibfield  {author} {\bibinfo {author} {\bibfnamefont {Y.}~\bibnamefont
  {Yao}}, \bibinfo {author} {\bibfnamefont {L.}~\bibnamefont {Ge}}, \bibinfo
  {author} {\bibfnamefont {X.}~\bibnamefont {Xiao}}, \bibinfo {author}
  {\bibfnamefont {X.-g.}\ \bibnamefont {Wang}},\ and\ \bibinfo {author}
  {\bibfnamefont {C.-p.}\ \bibnamefont {Sun}},\ }\href@noop {} {\bibfield
  {journal} {\bibinfo  {journal} {Phys. Rev. A}\ }\textbf {\bibinfo {volume}
  {90}},\ \bibinfo {pages} {022327} (\bibinfo {year} {2014})}\BibitemShut
  {NoStop}%
\bibitem [{Sup()}]{SupplementalMaterial}%
  \BibitemOpen
  \href@noop {} {}\bibinfo {note} {See Supplemental Material at [URL will be
  inserted by publisher] for the NumPy code used for the independent numerical
  checks.}\BibitemShut {Stop}%
\end{thebibliography}%
	
\end{document}